\documentclass[11pt]{article}
\usepackage{graphicx}
\usepackage[comma,square]{natbib}
\usepackage{textcomp}
\usepackage{amssymb,amsmath}

\begin{document}

\title{The Optimum Distance at which to Determine the Size of a Giant Air
  Shower}
\author{D. Newton, J. Knapp, A.A. Watson \\ School of Physics and Astronomy \\
  University of Leeds \\ Leeds LS2 9JT}
\date{\today}
\maketitle

\begin{abstract}

To determine the size of an extensive air shower it is not necessary to have
 knowledge of the function that describes the fall-off of signal size from the
 shower core (the lateral distribution function).
 In this paper an analysis with a  simple Monte Carlo
  model is used to show that an optimum ground parameter can be
  identified for each individual shower. At  this optimal core distance,
  $r_\mathrm{opt}$,  the fluctuations in the expected signal, $S(r_\mathrm{opt})$, due to a
  lack of knowledge of the lateral distribution function  are minimised.
  Furthermore it is shown that the optimum ground parameter is determined
  primarily by the array geometry, with little dependence on the energy or
  zenith angle of the shower or choice of lateral distribution function. For an array such as the
  Pierre Auger Southern Observatory, with detectors separated by 1500 m in a
  triangular configuration, the optimum distance at which to measure this
  characteristic signal is close to 1000 m. 

\end{abstract}

\newpage

\section{Introduction}

 The extreme rarity of the largest extensive air showers necessitates an
 observatory of large aperture if large numbers of events are to be recorded.
  In the case of an array of surface detectors, the need to cover as large an area
 as possible, combined with inevitable economical constraints, invariably leads to an array with  large
 separation between adjacent detectors. The result  is that properties of any individual air
 shower are sampled at a limited number of points at different distances
 from the shower core. When reconstructing the size of the shower, the lateral
 distribution function (LDF), which describes the fall-off of signal size with
 the distance from the shower core, must be assumed and any inaccuracies in it
 (either from uncertainties in the form of  the LDF or from intrinsic
 fluctuations in the development of the  air shower) will lead to a
 corresponding inaccuracy in both the location of the shower core and in the
 measurement of the integrated LDF (traditionally a measure of the total
 number of particles in the shower). To avoid the large fluctuations in the
 signal integrated over all distances, Hillas \cite{hillas, hillas1}  proposed using the
 signal at some distance from the shower core  to classify
 the size of the shower, and ultimately the energy of the primary particle.  In his
 original paper he pointed out  the practical advantages of this method: (i)
 the effect of uncertainties in the LDF are minimised at a particular core
 distance, and (ii) although the total number of particles at ground level is
 subject to large fluctuations, the fluctuations of the particle density far
 from the core are quite small. For example in ref.\cite{hillas} it is shown that
 at $10^{17}$ eV the RMS variation in the total number of particles is
 $\approx 67$ \%; for the same shower the RMS variation in the signal of a
 water-Cherenkov detector at 950 m is $\approx
 6$ \%. While these absolute numbers are certainly model dependent,
 his conclusion was shown to be robust for a variety of models and energies.

The measurement of the energy of a primary particle  with a surface array of
particle detectors, is thus  a two-step process. Firstly the detector signal
at a particular core distance must be measured and secondly this
characteristic signal must be linked to the energy of the primary particle. The choice of the distance at which to measure this  characteristic signal will depend on the combined (and independent) uncertainties of each step.  
All cosmic ray observatories which employ surface detectors (with the notable
exception of the Pierre Auger Observatory) have relied on models to perform
the second step. Systematic uncertainties aside, the effects of intrinsic
shower-to-shower fluctuations are minimised if the characteristic signal (the
shower ground parameter) is measured at a core distance of $\gtrsim 600$ m
\cite{hillas,hillas1}. 
The Pierre Auger Observatory \cite{eapaper}, with the benefit of its hybrid
 design, utilises fluorescence detectors to  measure around 10\% of the
 EASs observed with the surface array. The  calorimetric energy measurement
 from the fluorescence detectors can be used  to calibrate the
 characteristic signal from the surface detectors  \cite{sommers}, 
thus substantially reducing a large source of systematic uncertainty. 

The optimum core distance for  the first step, the measurement of a
characteristic signal, is solely dependent on the geometry of the
array. 
The optimum distance to measure the
shower ground parameter, S($r_\mathrm{opt}$), can be identified for each individual event. The work
of Hillas led to the Haverah Park Collaboration adopting $\rho (600)$, the
particle density at 600 m from the shower core measured in a 1.2 m deep water-Cherenkov detector, as the optimum energy estimator.

 Identifying this optimum core distance, $r_\mathrm{opt}$, is especially relevant to
giant arrays, where the large array spacing (typically $\gtrsim 1$ km) makes
measuring the LDF problematic, except in events of very high multiplicity. The low number of signals obtained from any one
event usually  makes measurement of the LDF on an event-by-event basis
impractical. Instead an average LDF, parameterised in measurable observables such as zenith angle and energy, must be used. Even if the average LDF can be measured to a high degree of accuracy (which may not always be possible with a surface detector alone), intrinsic shower-to-shower fluctuations will affect the lateral structure of any particular event, often characterised as the slope of the LDF, describing how rapidly the particle density decreases with core distance. Identifying the optimum core distance at which to measure the characteristic signal of an air shower, will minimise the uncertainty in this parameter due to a lack of knowledge of the true LDF.

In this paper, a method of identifying this optimum `ground parameter', where the observed signal shows the least dependence on the assumed LDF,  is discussed and its dependence on energy, zenith angle  and array geometry is investigated.

\section{Simple Monte Carlo Analysis}\label{tmc}

The signals observed by an array of surface detectors were simulated using a simple Monte Carlo technique. This was achieved by adopting a lateral distribution function  for an extensive air shower:
$$ S(r) = k\, f(\beta, r) \, ,  $$
where $S(r)$ is the signal observed at a core distance $r$, $k$ is a size
parameter, and $f(\beta, r)$ is the functional form of the LDF, parameterised
as a function of the slope parameter, $\beta$, and the core distance,
$r$. Throughout this paper, the signal is expressed as that relative to a
vertical equivalent muon (VEM), although any arbitrary unit would give a
similar result. The simulated shower was  projected onto a virtual array with
a specified core location and arrival direction and the signals in each tank,
at the calculated core distance, were given by the assumed LDF. The signals
were then varied according to a Poissonian distribution in the number of
particles, assuming each particle contributes 1 VEM. Signals below a threshold of 3.2 VEM were set to zero, and
those above 1000 VEM were treated as saturated. These values were chosen to accord with what is
currently used at the Auger Observatory. Using the known arrival direction as an input, the event can then be analysed to assess the reconstruction algorithm, in particular the effect of using an incorrect LDF.

\begin{figure}[htp!]
\centerline{
\includegraphics[width=0.45\textwidth]{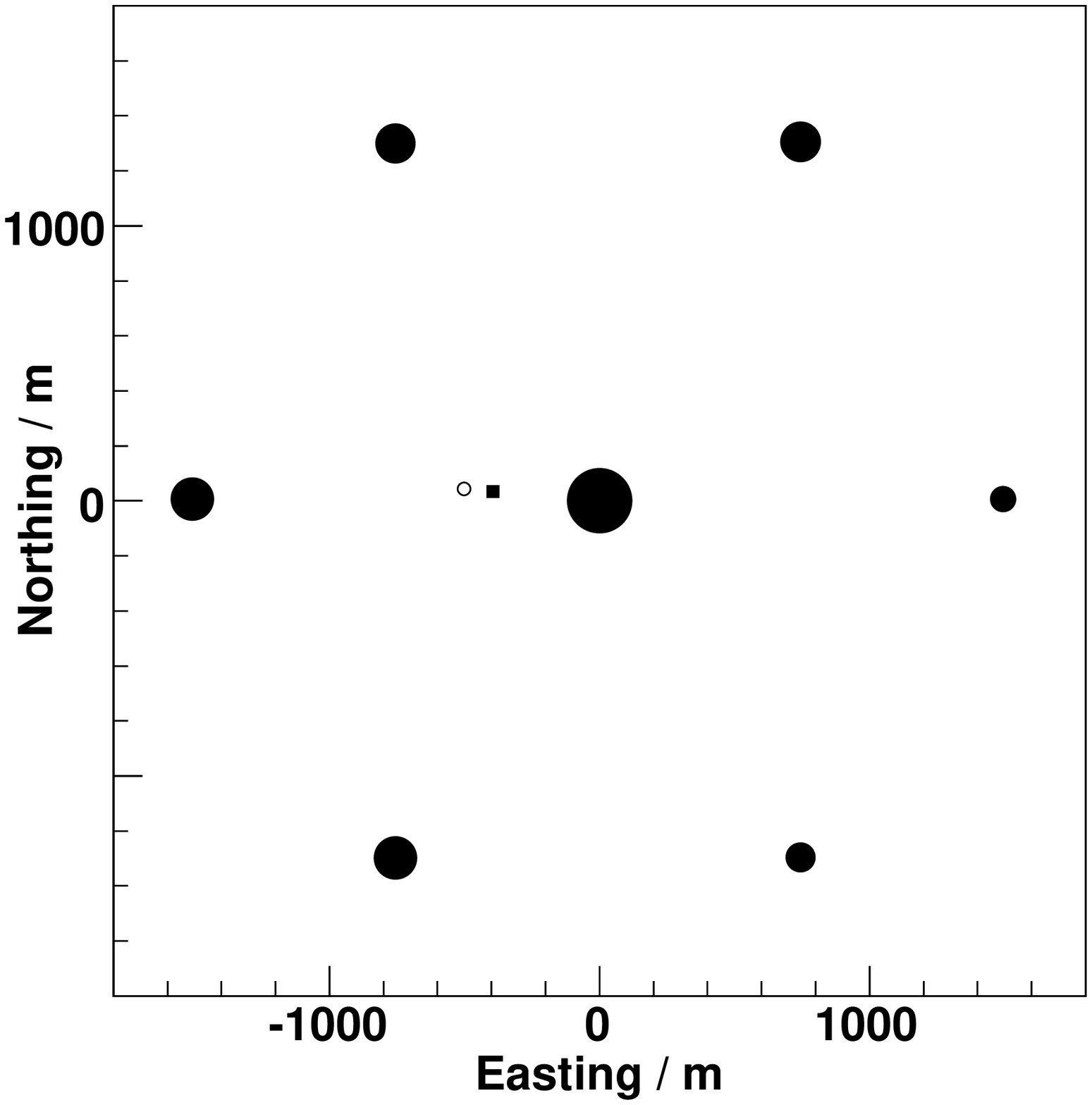}}
\centerline{
\includegraphics[width=0.75\textwidth]{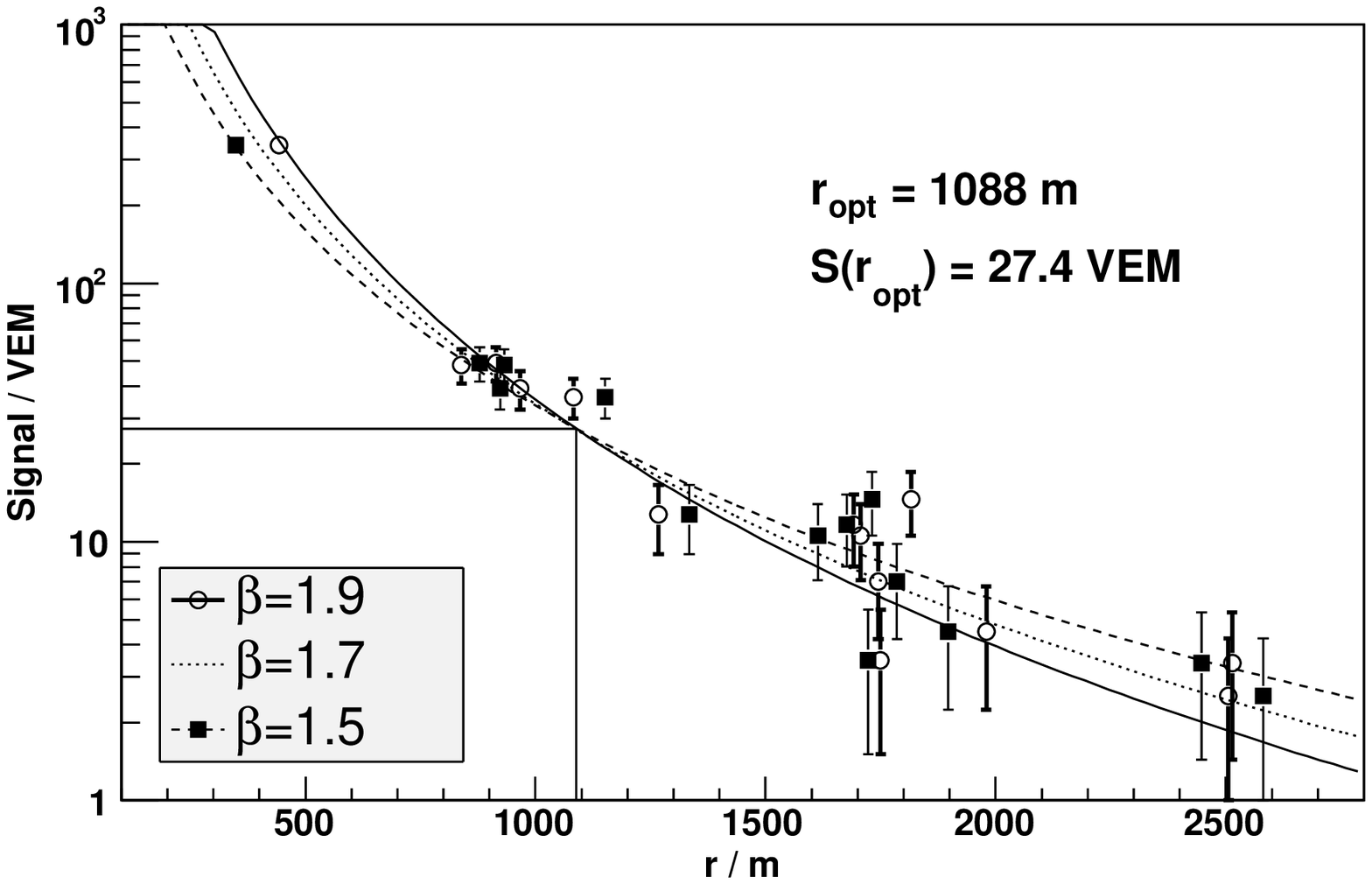}}
\caption{\it The optimum core distance for one event. The same event was
  reconstructed with an `NKG' type LDF, using two different values for the
  slope parameter: $\beta=1.5$ (filled squares), $\beta=1.7$ (dotted line),  and $\beta=1.9$ (circles). The  different slope parameters result in different reconstructed core locations
  (separated by $\sim 100$ m), indicated by the open circle and the square on the plan
  of part of the array (top) where the black points indicate tanks with a signal
  and the size of the point is proportional to the logarithm of the signal. Both reconstructions give reasonable fits to the signals (bottom).  The
  dotted line shows the reconstruction using an intermediate value of
  $\beta=1.7$ (points not plotted). At  $\sim r_\mathrm{opt}$, the same signal ($\sim
  27$ VEM) is measured for each reconstruction, and by converting this ground parameter, $S(r_\mathrm{opt})$ into the energy of the primary particle, uncertainties due to a lack of knowledge of the true slope parameter are minimised.}\label{ldfeg}\end{figure}

A hexagonal  array was chosen with detector spacing of 1500 m corresponding
to that of the Pierre Auger Southern Array.  The dependence of the event reconstruction on the slope parameter of the LDF is shown in figure \ref{ldfeg}. For this single event the input LDF was an `NKG' type function with the form 
$$ S = k \left ( \frac{r}{r_s} \right ) ^{-\beta}  \left (1+ \frac{r}{r_s} \right ) ^ {-\beta}\, . $$

The slope parameter $\beta$ was parameterised as a function of zenith angle, $\theta$, such that $ \beta = 2.5 - (\sec \theta -1) $
, the scaling parameter, $r_s$ was set to 700 m and the size parameter, $k$,
was set to 250 (comparisons with full Monte Carlo simulations suggest this
corresponds to  a shower initiated by a primary with  energy of $\sim 10$
EeV). The zenith angle was set to $55^\circ$. From this LDF signals for each
tank were drawn and varied according to typical (Poissonian) measurement
uncertainties, as described above. To reconstruct the shower, the
`NKG' type function was used to fit the location of the shower core and the
size parameter, $k$, simultaneously. The reconstruction of the event was
carried out three times using different fixed values for the slope parameter
$\beta$ with each reconstruction. The effect of the differing values for  the slope
of the LDF is clearly seen in figure \ref{ldfeg}. While the fits result in a
substantial shift in the reconstructed core location (of $\sim 100$ m) and 
corresponding shifts in the core distances of the individual detectors, it is
evident that the characteristic signal for this event, at $\sim 1090$ m, is
very similar for all three reconstructions. Measuring the signal at this point minimises the effect of the systematic uncertainty in the slope parameter of the LDF.

\section{Determining $r_\mathrm{opt}$}

To find the distance, $r_\mathrm{opt}$, for which the signal variation with respect to the slope parameter, $\beta$, is smallest, one can minimise $\frac{dS}{d\beta}$. For a power law LDF, this is shown in the following. 

$$ S = kr^{-\beta} = ke^{-\beta \ln r} \ , $$

 where $S$ is the predicted signal, k is the shower size parameter and $\beta$ is the slope parameter.

If $k = k(\beta) $, 
\begin{equation}
 \frac{\mathrm{d}S}{\mathrm{d}\beta} = \frac{\mathrm{d}k}{\mathrm{d}\beta} e^{-\beta \ln r} + k e^{-\beta \ln r}(-\ln r) 
\label{sys}
\end{equation}

\centerline{$ \frac{\mathrm{d}S}{\mathrm{d}\beta} = e^{-\beta \ln r} \left (\frac{\mathrm{d}k}{\mathrm{d}\beta} - k \ln r \right )= 0$, at $r_\mathrm{opt}$ }
\noindent so,
$$ \frac{\mathrm{d}k}{\mathrm{d}\beta} = k \ln r_\mathrm{opt}$$

and, 
\begin{equation}
\frac{\mathrm{d} \ln k}{\mathrm{d}\beta} = \ln r_\mathrm{opt}
\label{ropt}
\end{equation}

A similar deduction  can be applied to different classes of  LDF, such as a `Haverah Park' function: 

$$ S = k  r^{-(\beta + \frac{r}{4000})}$$

$$ \Rightarrow \frac{ \mathrm{d} (\ln k)}{\mathrm{d} \beta} = \ln r_\mathrm{opt} \, ,  $$

 or an `NKG' type function:

$$ S = k \left ( \frac{r}{r_s} \right ) ^{-\beta}  \left (1+ \frac{r}{r_s} \right ) ^ {-\beta} $$

\begin{equation}
 \Rightarrow \frac{ \mathrm{d} (\ln k)}{\mathrm{d} \beta} = \ln \left ( \frac{r_\mathrm{opt}}{r_s} \right ) + \ln \left (1+ \frac{r_\mathrm{opt}}{r_s} \right ) 
\label{roptnkg}
\end{equation}

\noindent which leaves a quadratic equation to be solved to find $r_\mathrm{opt}$:
\begin{equation}
\frac{r_\mathrm{opt}}{r_s}=\frac{-1+\sqrt{1+4e^{\alpha}}}{2}
\end{equation}
where $\alpha=\frac{ \mathrm{d} (\ln k)}{\mathrm{d} \beta}$.

Of course, in a real event, the observed signals are subject to a measurement
uncertainty and the LDF must be fitted to the data using a suitable
minimisation procedure, but  using reasonable values for  $\beta$ and the size
of the fluctuations gives a good approximation to this analytical
solution. $r_\mathrm{opt}$ can then be found for any event by analysing it several
times, using different values for the slope parameter and  either plotting
$ln\, k$ against $\beta$, or numerically minimising the spread of S(r) at any
core distance, $\Delta S(r)$.

\begin{figure}[t!]
\centerline{
\includegraphics[width=\textwidth]{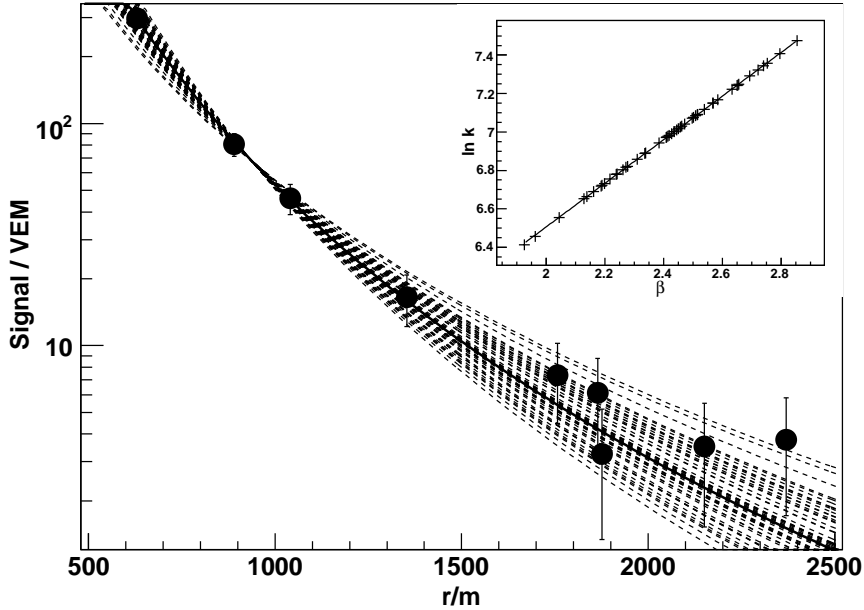}}
\caption{\it Reconstructing the same event 50 times, using different values
  for the slope parameter, $\beta$, allows the optimum ground parameter,
  $r_\mathrm{opt}$ to be found (using an `NKG' type LDF in this example). At $r_\mathrm{opt}$,  the expected signal is essentially
  independent of the slope parameter. The inset plot shows the  relationship between the slope parameter and the fitted size parameter,
  allowing $r_\mathrm{opt}$ to be   calculated analytically (equation \ref{roptnkg}).}\label{showropt}
\end{figure}

A simulated event is shown in figure \ref{showropt}  after reconstructing the
event 50 times, using values of the slope parameter, $\beta$, drawn from a
Gaussian distribution with a mean of 2.4 and a width of 10\%, which
corresponds approximately to the uncertainty in $\beta$. The zenith angle
of the event is $24^\circ$ and the size parameter, $k$, was set to 1050
(again, corresponding to a primary energy of $\sim 10$ EeV). The value taken for
the magnitude of the intrinsic fluctuations in the slope parameter is based on
measurements made at Haverah Park \cite{ave,coy}, which indicate that  10\% is
an appropriate  value. Fluctuations of a similar magnitude were measured at
Volcano Ranch \cite{vr}. The reconstructed LDFs can be seen to converge at around 940 m. The inset panel shows $\beta \,vs \, \ln k$ found from the 50 reconstructions. The relationship is approximately linear and, using the formula given in equation \ref{roptnkg},  the spread in $S(r)$ is found to be minimised at 938 m. The spread, $\Delta S(r)$, at any core distance corresponds to the systematic uncertainty in $ S(r)$ due to the uncertainty in the slope parameter, $\beta $ and can be found analytically by equating $$ \frac{d S(r)}{d \beta} \approx \frac{\Delta S(r)_\mathrm{sys}}{\Delta \beta} $$ and rearranging equation \ref{sys} for the appropriate LDF.

\begin{figure}[t!]
\centerline{
\includegraphics[width=\textwidth]{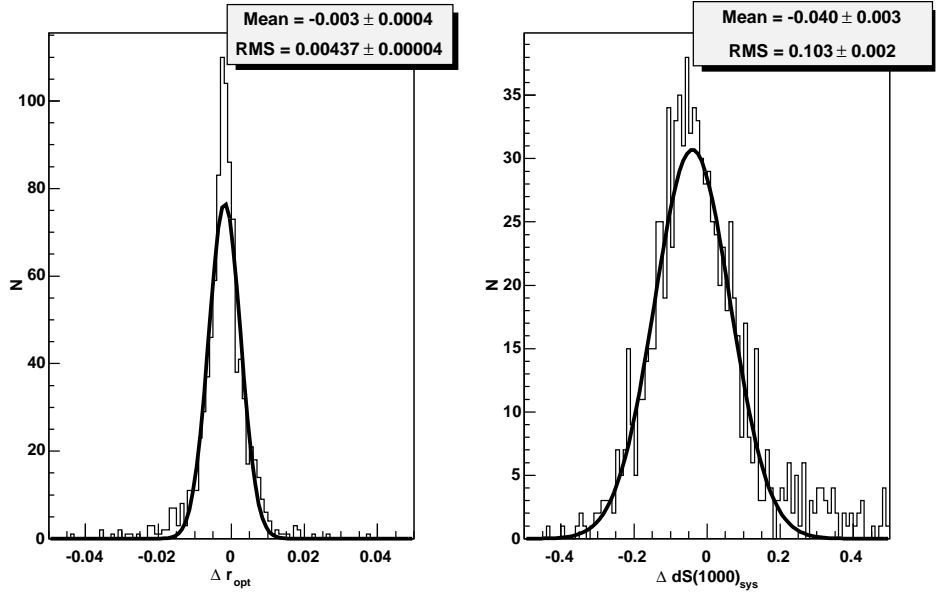}}
\caption{\it Comparing the optimum ground parameter (left) and the systematic
  uncertainty in S(1000) (right) found using a numerical and an analytical
  technique. In each case the residual is shown as
  $\frac{numerical-analytical}{numerical}$. The statistics refer to the
  Gaussian fit.}\label{num_ana_comp}
\end{figure}

A comparison of the analytical and numerical techniques, found from 1000
simulated showers is shown in figure \ref{num_ana_comp}. Each shower had a
randomly assigned zenith angle, $\theta$, of between 0 and  $60 ^\circ$
following a flat distribution in $\sec \theta$. The size parameter, $k$, was
also randomly assigned from a flat distribution, to give a range of equivalent
energies of between 5 and 100 EeV. The slope parameter, $\beta$ for each event
had the  mean value found using the parameterisation given in section
\ref{tmc}, but was then varied according to a Gaussian distribution with a
width of 10 \%. The resulting values of  $r_\mathrm{opt}$ and $\Delta S(1000)_\mathrm{sys}$ agree within 1\% and 10 \% respectively, demonstrating that the analytical technique is a good approximation to the numerical analysis.

\subsection{Saturated Signals}

\begin{figure}[th]
\centerline{
\includegraphics[width=\textwidth]{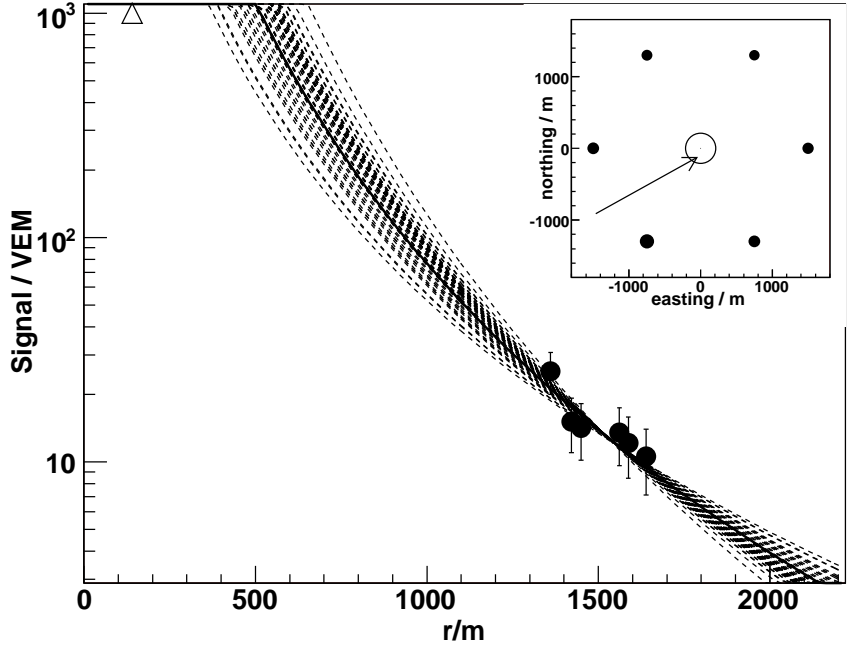}}
\caption {\it $ r_\mathrm{opt}$ for a saturated, vertical ($\theta=24^\circ$)  event. With 6 tanks at roughly the
  same distance ($\sim 1500$ m), this is the optimum point to measure the
  shower ground parameter. The open triangular point indicates the saturated
  tank. The inset plot shows a plan of the array - the saturated   tank (open
  circle) is surrounded by a hexagon of tanks with signals (filled
  circles). The shower core and azimuthal direction are indicated by the arrow
  .}\label{satevent}
\end{figure}

If, by chance, a shower core lies  close to one tank, the signal in that
tank may be saturated: for the Auger Observatory, currently, saturation occurs
for signals larger than $\approx$ 1000 VEM. The effect of a saturated signal on the
reconstruction algorithm can be seen in  figure \ref{satevent}.  This vertical
shower  has one saturated tank, then a ring of 6 tanks with similar signals, all lying at about 1500 m from the core. The reconstructed values are only  weakly constrained by the presence of the saturated tank, and so $r_\mathrm{opt}$ is found at the point where most of the signals are measured ($\sim 1500$ m).

\subsection{Dependence on Zenith Angle and Energy}

\begin{figure}[ht!]
\centerline{
\includegraphics[width=0.5\textwidth]{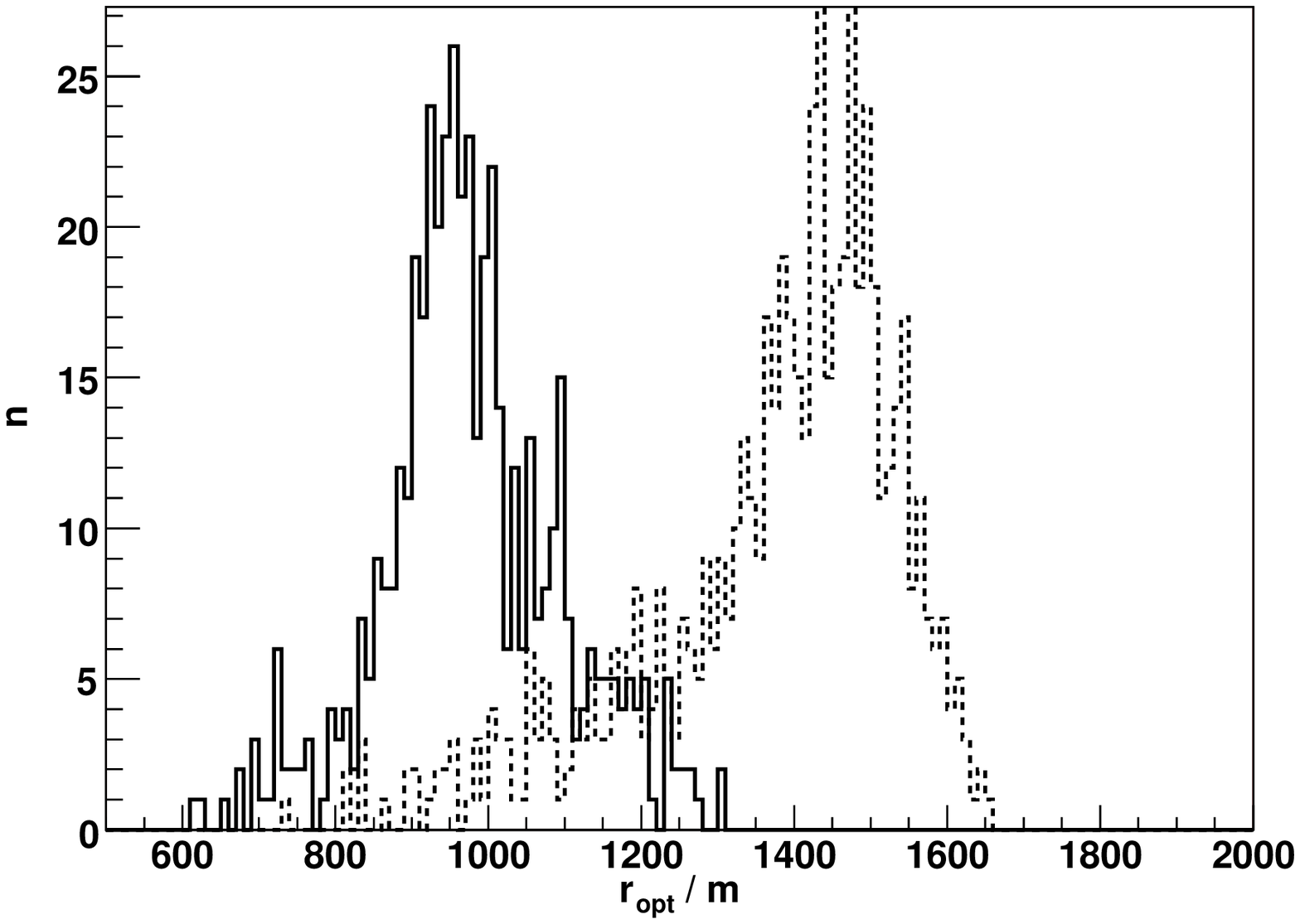}
\includegraphics[width=0.5\textwidth]{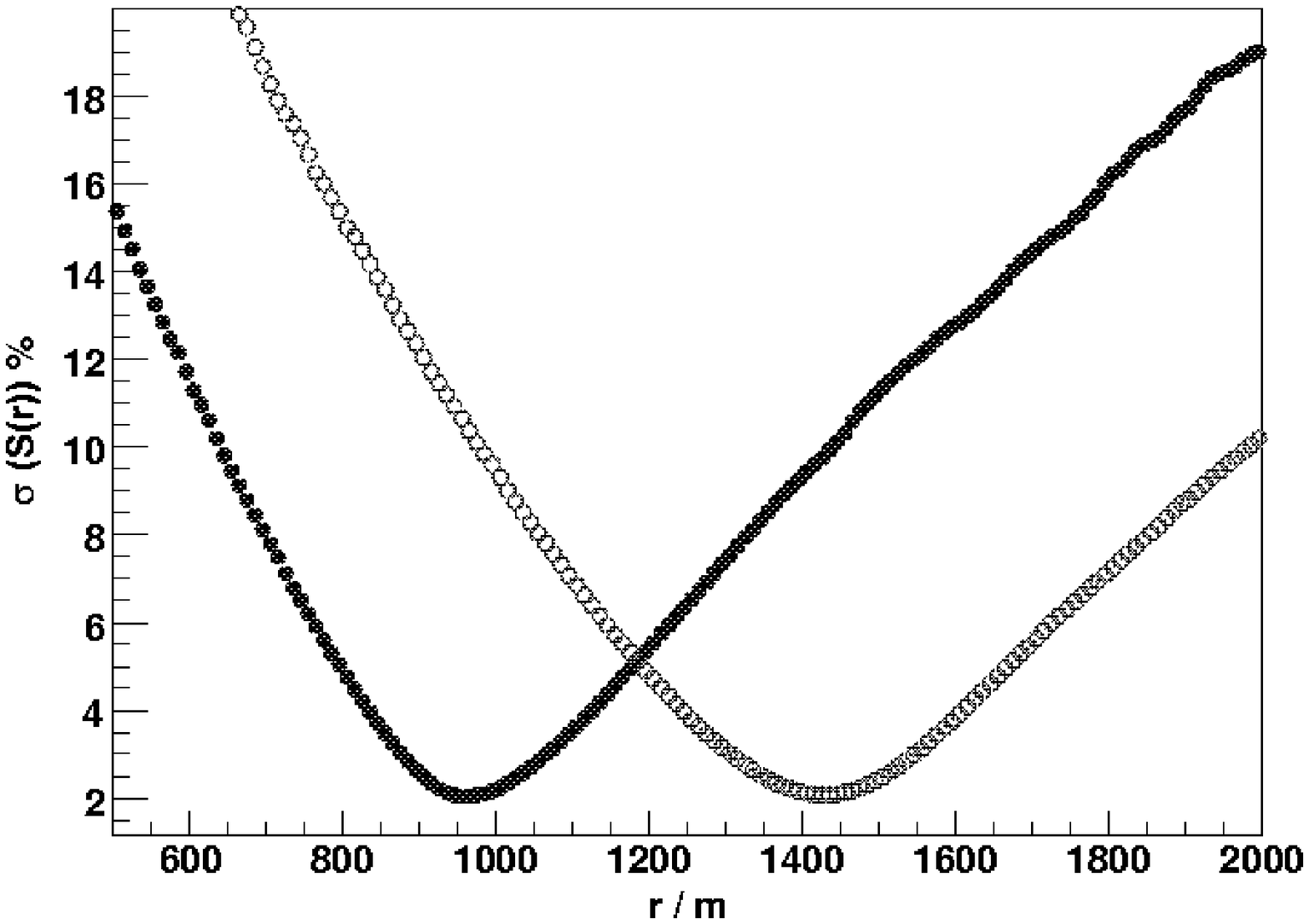}}
\centerline{
\includegraphics[width=0.5\textwidth]{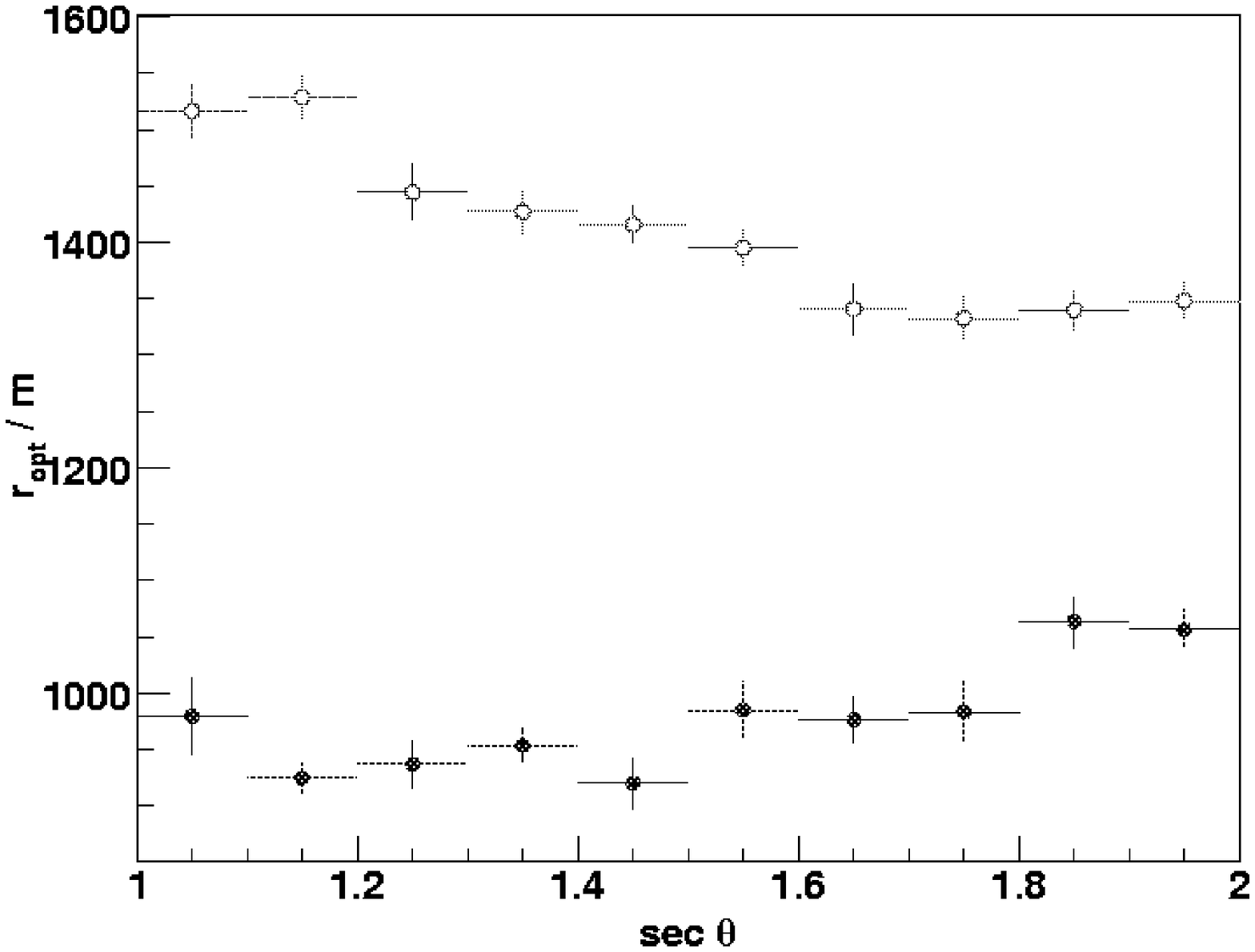}
\includegraphics[width=0.5\textwidth]{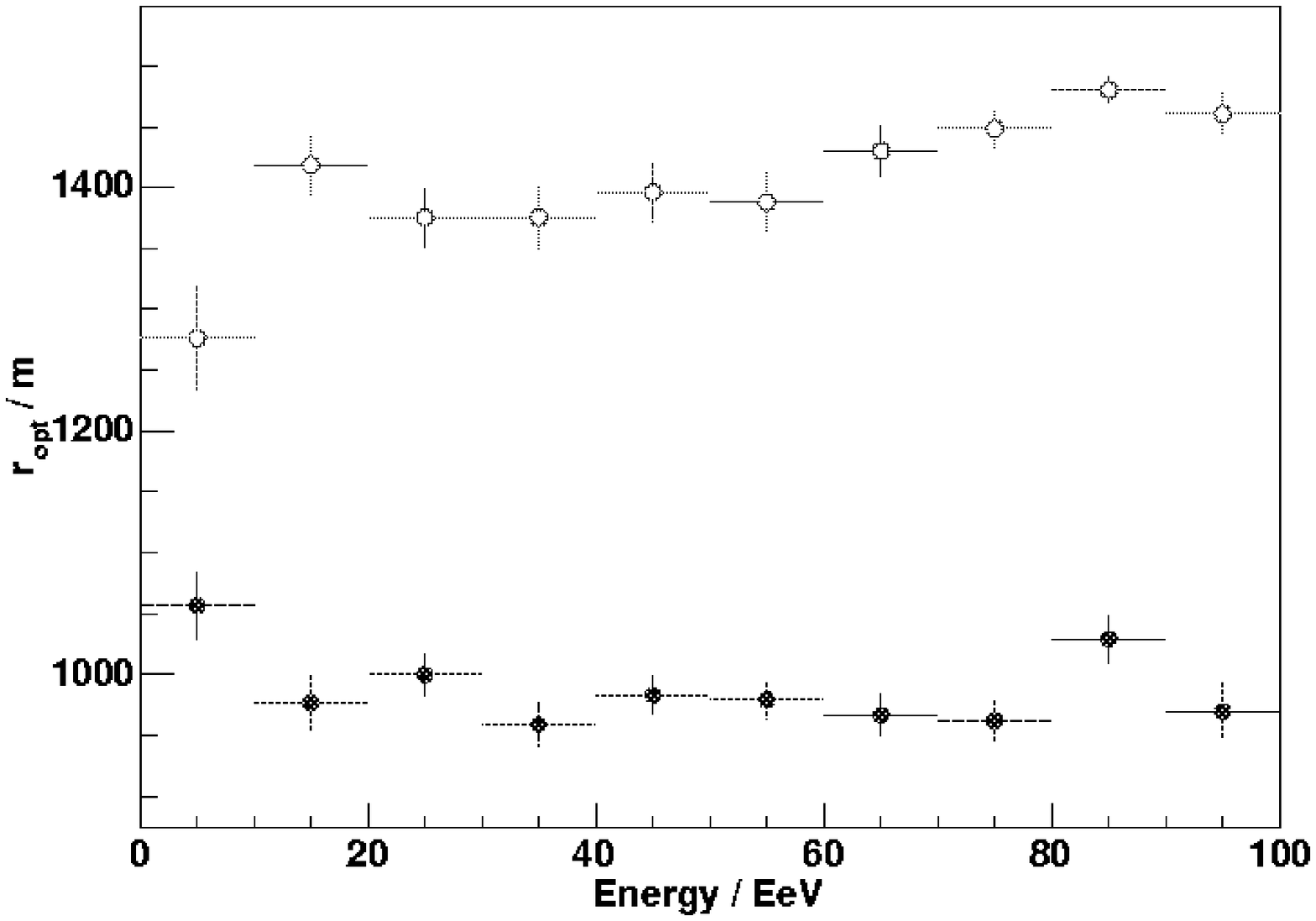}}

\caption{\it Top left panel: The distribution of $ r_\mathrm{opt}$ (assuming $10\%$
  fluctuations in the slope parameter), for $10³$   simulated showers, with
  zenith angle  $ 1.0<\sec \theta < 2.0$ and energies between 5 and 100
  EeV. The dashed line shows the distribution of 
  saturated signal while the solid line is for events without a saturated
  signal. Top right panel: The mean percentage spread in $S(r)$.  Bottom row:
  The dependence of $r_\mathrm{opt}$ on the zenith angle, $\theta$, and energy of the
  primary particle. In all  plots, the solid  markers indicate events in which
  there are no saturated signals while  the open markers  are for events with one or more saturated signals.}\label{roptfluc}
\end{figure}

An analysis of 1000 showers with  zenith angles between 0 and $60^\circ$ and
size parameters such that $5 < E < 100$ EeV gives the  distribution of
$r_\mathrm{opt}$ shown in figure \ref{roptfluc} (top left panel). The distribution
has two distinct populations corresponding to events with and without
saturated signals. The mean value for the showers without a saturated signal
is $970 \pm 1$ m and the RMS deviation is $102 \pm 1$ m. The mean spread in
$S(r)$ as a function of $r$ is shown in the upper right panel of the same
figure. If the characteristic signal size is measured at 970 m for all the
showers, the mean systematic uncertainty in this measurement is less than 2\%,
although this increases to $\sim 10$ \% for the events with a saturated
signal. The relatively small change of $r_\mathrm{opt}$ with zenith angle is clear
in the bottom left hand diagram of the figure.

If the same events are reconstructed with  different forms of
LDF, such as a power law or a `Haverah Park' type LDF, the optimum ground
parameters can  be found with the same technique and give  similar
results as those found using the `NKG' type function. The distribution
of $r_\mathrm{opt}$, for the events without a saturated signal,  found using 3
different LDFs are shown in table \ref{table2}. Also shown in table
\ref{table2} are the values of S(1000) measured with each LDF, normalised to
S(1000) measured with the `NKG' type  LDF. The agreement between the values of
S(1000) found using these 3 different LDFs is better than 5\%. 

\begin{table}[b]
\begin{center}
\begin{tabular}{|c|c|c|c|}
\hline
LDF    & $ r_\mathrm{opt}$ / m, mean & $ r_\mathrm{opt}$ / m, RMS & $\Delta S(1000)=
\frac{S(1000)_\mathrm{LDF}}{S(1000)_\mathrm{NKG}} $ \\ \hline \hline
Power Law      & 960  &  110 & $1.045 \pm 0.001$  \\ \hline
`Haverah Park' & 940  &  100 & $0.986  \pm 0.001$ \\ \hline
`NKG' type      & 970  &  110 & 1.00 \\ \hline
\end{tabular}
\caption{\it The optimum ground parameter found using 3 different LDFs}\label{table2}
\end{center}
\end{table}

\begin{figure}[ht!]
\centerline{
\includegraphics[width=9.0cm,height=6.0cm]{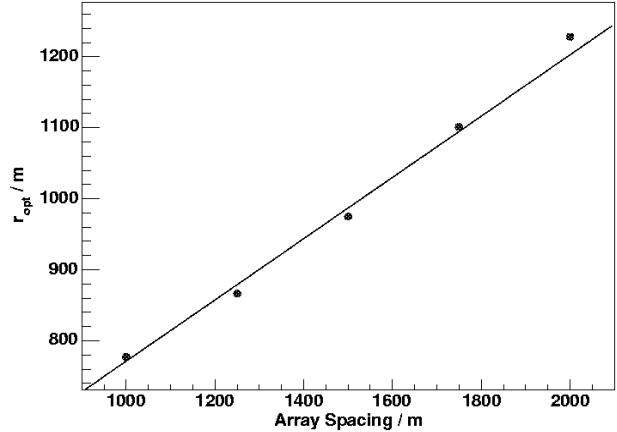}}
\caption{\it $r_\mathrm{opt}$ as a function of the surface array spacing. The
  uncertainties in $r_\mathrm{opt}$ are smaller than the points.}\label{aspace}
\end{figure}

The variation of $r_\mathrm{opt}$ as a function of the array spacing is shown in figure \ref{aspace}. The relationship is approximately linear with the value of $r_\mathrm{opt}$  increasing at a rate of $\sim 45$ m for each 100 m increase in the detector separation. 

\section{Conclusion}

An analysis of the  events simulated with a simple Monte Carlo model has shown that the optimum core
distance to measure the size of the shower can be calculated for
each shower and is determined primarily by the array geometry, with no
significant dependence on the shower zenith angle, energy or the assumed
lateral distribution function (figure \ref{roptfluc}, table \ref{table2}). The
presence of saturated tanks has a significant impact on $ r_\mathrm{opt}$, resulting
in an increase of up to 500 m (depending on the zenith angle).  This will primarily affect large,
vertical showers that have a relatively small number of tanks with signals and
a high probability of saturation in one tank. The determination of $r_\mathrm{opt}$ for these events will depend on the treatment of these signals in the reconstruction algorithm. 

For an array with 1500 m spacing, such as the Pierre Auger Observatory,  the
plot shown in figure \ref{aspace} indicates that $r=1000$ m is a good choice at
which to measure the characteristic signal, S(1000), used to determine the
energy of the primary particle. At around 1000 m the expected signal is robust
against inaccuracies in the assumed LDF at better than 5\%.

\section{Acknowledgments}
Research on Ultra High Energy Cosmic Rays at the University of
Leeds is supported by PPARC, UK.  DN is grateful for the support
of a PPARC Research Studentship.

\bibliographystyle{ieeetr}

\end{document}